\documentclass[%
 reprint,
 amsmath,amssymb,
 aps,
]{revtex4-1}
\usepackage{afterpage}
\usepackage{color}
\usepackage{graphicx}% Include figure files
\usepackage{dcolumn}% Align table columns on decimal point
\usepackage{bm}% bold math
\newcommand{\comment}[1]{}

\begin{document}

\preprint{APS/123-QED}

\title{Nutation Wave as a Platform for Ultrafast Spin Dynamics in Ferromagnets}% Force line breaks with \\

\author{Imam Makhfudz$^{\psi}$ Enrick Olive$^{\psi}$ and Stam Nicolis$^\phi$}
\affiliation{%
%\centering
$^{\psi}$GREMAN, UMR 7347, Universit\'{e} de Tours-CNRS, INSA Centre Val de Loire, Parc de Grandmont, 37200 Tours, France\\ 
$^{\phi}$Institut Denis Poisson, Universit\'{e} de Tours, Universit\'{e} d'Orl\'{e}ans, CNRS (UMR7013), Parc de Grandmont, F-37200, Tours, France
}%
\date{\today}% It is always \today, today,

\begin{abstract}
At short time scales the  inertia term becomes relevant  for  the magnetization dynamics of ferromagnets  and leads to nutation for  the magnetization vector. For the case of spatially extended magnetic systems, for instance Heisenberg spin chains with  isotropic spin-exchange interaction, this leads to  the appearance of a collective excitation, the ``nutation wave", whose properties are elucidated by analytic arguments and numerical studies. The one--particle excitations can be identified as relativistic massive particles.  These particles,  the ``nutatons", acquire their mass via the Brout-Englert-Higgs mechanism, through the interaction of the wave with  an emergent topological gauge field. This spin excitation would appear as a peak in the spectrum of the scattering structure factor in  inelastic neutron scattering experiments. The   high frequency and speed of the nutation wave can  open paths for realizing ultrafast spin dynamics.

%\begin{description}



%\end{description}
\end{abstract}

\pacs{Valid PACS appear here}% PACS, the Physics and Astronomy
\maketitle

Spin dynamics plays a  crucial role in technological applications of magnetism and is one of the mainstays in the emerging field of spintronics \cite{RMPspintronics}. As the demand for efficient and high speed electronic devices is becoming more urgent than ever, the drive towards  fast or even ultrafast spin dynamics \cite{UltrafastReview1}\cite{UltrafastReview2}\cite{RMPultrafast} has been an object of intense pursuit by the research community. Of particular importance is the magnetic switching, which in most electronic devices relies on domain wall motion and precessional motion of the magnetization vector, both of which are however relatively slow and rely on very strong magnetic fields. This  limits the speed of the available magnetic data storage devices \cite{Ultimate}.

Precessional motion and spin wave  propagation describe  low-energy, massless, excitations of spin systems. While the spin dynamics is described by the well known Landau-Lifshitz-Gilbert equation \cite{LL}\cite{GilbertIEEE}, recent theoretical studies using  complementary  approaches have found that a term, corresponding to the inertia of a spin or single domain of uniformly magnetized ferromagnet, appears at short time scales~\cite{Ciornei}\cite{AJP}\cite{APL}\cite{JAP}\cite{FermiSurfaceBreathingModel}\cite{Atomistic} and is also distinct from the inertia of topological defects or spin textures, such as that of a Skyrmion bubble \cite{MakhfudzPRL2012}. A useful mechanical analogy is the spinning top:  This inertia term originates from the transverse components of the  inertia tensor $I_{xx(yy)}$~\cite{Ciornei}\cite{AJP}. It gives rise to a so far-neglected type of spin motion;  the nutation of the spin, which is a cycloidal motion with a high characteristic frequency, which is of particular relevance in  ultrafast spin dynamics. A recent experiment has reported the observation of such high frequency nutation dynamics \cite{Neeraj}.

In this Letter, we point out that a type of collective excitation appears in spin chains, in the presence of such a one-site inertia term, when the spins interact through  a spin exchange interaction. This excitation takes the form of a wave of nutation, to be referred to as the  ``nutation wave". The corresponding single-particle excitation, called the ``nutaton" here, is found to have a gap at zero wave vector and the dispersion relation  of a relativistic particle, that we identify with  the Higgs mode arising from the coupling between the scalar bosons, that describe the magnetization profile and a emergent gauge  boson, that describes the phase of the profile, along the chain. Experiments that could probe these properties are proposed. 
  
We can describe  the appearance of the nutation wave as follows. Consider the single--site  inertial Landau-Lifshitz-Gilbert (ILLG) equation \cite{Ciornei}\cite{AJP}\cite{APL}\cite{JAP}:
\begin{equation}\label{ILLG}
\frac{d\mathbf{M}}{dt}=\gamma \mathbf{M}\times\left[\mathbf{H}_{\mathrm{eff}}-\eta\left(\frac{d\mathbf{M}}{dt}+\tau\frac{d^2\mathbf{M}}{dt^2}\right)\right]
\end{equation}
in terms of the magnetization vector $\mathbf{M}$, where $\gamma$ is the gyromagnetic constant, $\eta$ is the Gilbert damping and $\tau$ is the spin relaxation time. The equation (\ref{ILLG}) has a microscopic origin as a torque equation and is applicable even to an individual spin (or magnetic moment). The second time derivative is an inertia term that has been shown to give rise to nutation, that describes oscillatory motion on top of the damped precessional motion produced by the first two terms on the right hand side \cite{Ciornei}\cite{AJP}\cite{APL}\cite{JAP}. 

Existing studies have so far assumed perfectly uniform magnetization, subject to uniform external field $\mathbf{H}_{\mathrm{eff}}=\mathbf{H}$. In realistic situations, there would be a spatial variation in the magnetization even in single domain system. We focus here on the consequences of such a spatial variation. In the presence of exchange interaction, the effective field is given by $ \mathbf{H}_{\mathrm{eff}}=\mathbf{H}+J_{ij}\partial^2\mathbf{M}/\partial x_i\partial x_j$ \cite{LandauLifshitzStatPhys2} in the ILLG equation eq.(\ref{ILLG}), which, on considering a one-dimensional isotropic Heisenberg model for which $J_{ij}=J\delta_{ij},i,j=x$, now becomes
\begin{equation}\label{ILLGspatial0}
\frac{d\mathbf{M}}{d t}=\gamma \mathbf{M}\times\left[\mathbf{H}-\eta\left(\frac{d \mathbf{M}}{d t}+\tau\left(\frac{d^2\mathbf{M}}{d t^2}-v^2_{\mathrm{iner}}\frac{\partial^2\mathbf{M}}{\partial x^2}\right)\right)\right]
\end{equation}
where 
\begin{equation}\label{wavevelocity}
v^2_{\mathrm{iner}}=\frac{J}{\eta\tau}
\end{equation}
Eq.~(\ref{ILLGspatial0})  is a  nonlinear wave equation. We shall study its solutions for the case of a  a single ferromagnetic domain with small spatial non-uniformities, i.e. no domain walls or solitons. 

In order to describe small fluctuations or excitations, it is necessary to define an appropriate background and study the linear excitations about it. To this end we shall  linearize eq.~(\ref{ILLGspatial0}), by writing its solution in the form
\begin{equation}\label{expanding}
\mathbf{M}(x,t)=\mathbf{M}_0+\widetilde{\mathbf{M}}(x,t)
\end{equation}
where  $\mathbf{M}_0$ is a spatially uniform magnetization. Furthermore, we rewrite eq.(\ref{ILLGspatial0}) in space-time translational invariant form that give the following equation
\[
D_g \widetilde{\mathbf{M}}=
\]
\begin{equation}\label{ILLGspatial}
\gamma \left[\mathbf{M}_0\times\left(\tilde{\mathbf{H}}-\eta\left( D_g+ \tau' D_{s=+}D_{s=-}\right) \tilde{\mathbf{M}}\right)-\tilde{\mathbf{H}}\times\tilde{\mathbf{M}}\right]
\end{equation}
where the space-time translation operator is given by $D_{v}=\partial_{t'}-v\partial_{x'}$ (with $v=\pm\tilde{v}_g,\tilde{v}_{s=\pm}$). The velocity $\tilde{v}_g$ is the characteristic velocity of a travelling wave solution of the eq.(\ref{ILLGspatial}) while $\tilde{v}_{s=\pm}=s v_{\mathrm{iner}}\pm v_g$ is another velocity that reflects the relativistic part of the equation of motion. eq. (\ref{ILLGspatial}) contains a  part that mimics the one in the standard wave equation \cite{WikipageAlembert}, 
\begin{equation}\label{simple1dwaveequation}
D^2_c\psi=\left(\frac{\partial}{\partial t}-c\frac{\partial}{\partial x}\right)\left(\frac{\partial}{\partial t}+c\frac{\partial }{\partial x}\right)\psi=0
\end{equation}
describing a wave  $\psi(x\pm ct)$, traveling with speed $v_{s=\pm}=s c$.  Recent works \cite{Ciornei}\cite{AJP}\cite{APL}\cite{JAP} have noted that  the inertia term gives rise to a cycloidal motion on top of the precessional motion, called nutation.  Its propagation along the chain, we thereby refer to  as a ``nutation wave", in direct analogy to the spin wave, which is a traveling wave of precession, with $v_{\mathrm{iner}}$ as the ``characteristic velocity" of the former. The principal difference between nutation and precession is that the former changes $M_{\perp}=\sqrt{M^2_y+M^2_z}$ while the latter does not; both cases however conserve $|\mathbf{M}|=\sqrt{M^2_x+M^2_{\perp}}=M_s$. 

\begin{figure}
\includegraphics[angle=0,origin=c, scale=0.20]{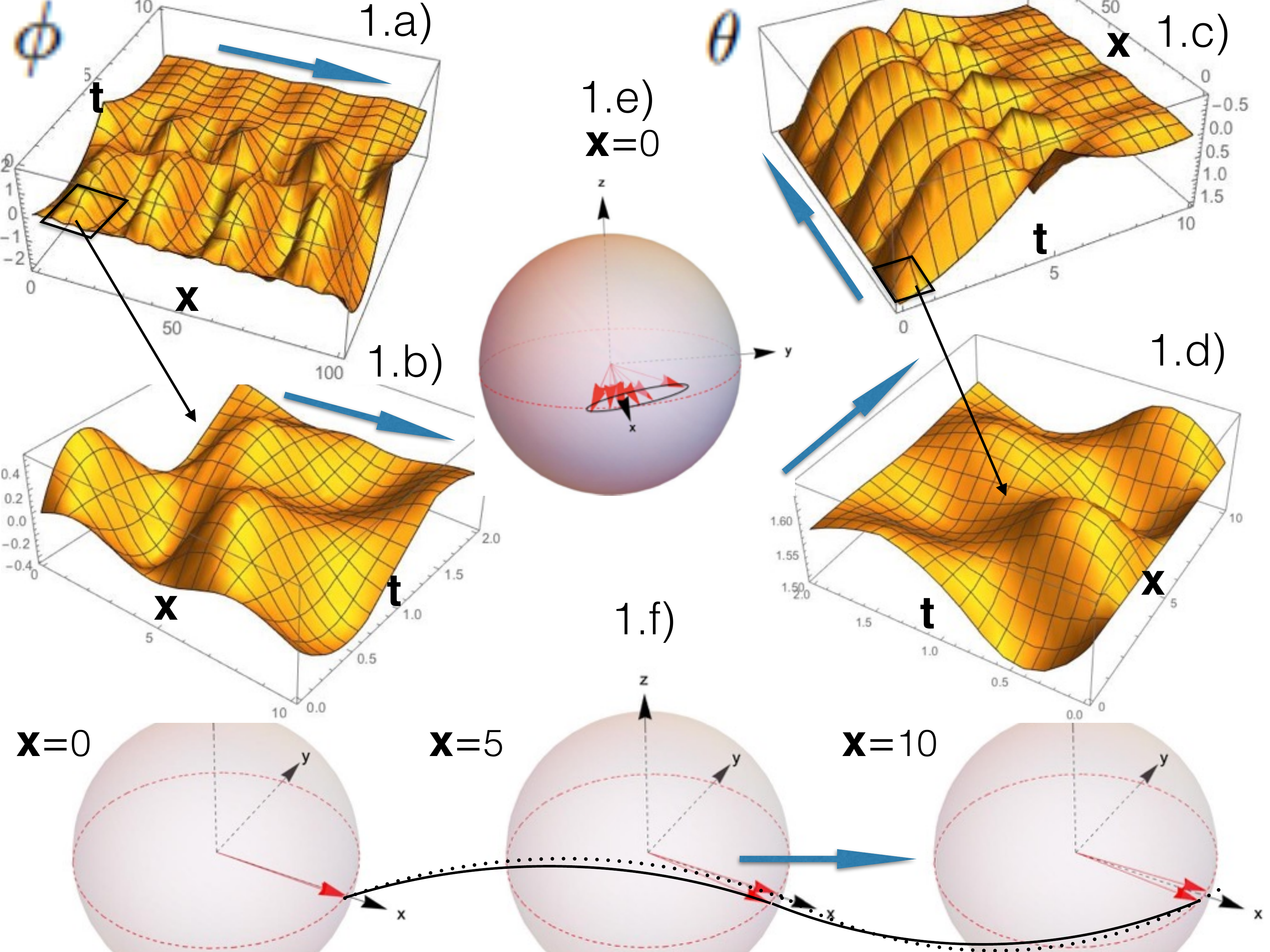}
 \label{fig:NutatonWaveProfile}
 \caption{
  The time evolution of the spins in terms of the spherical angles: $\phi(x,t)$( a) and b)) and $\theta(x,t)$ ( c) and d)) (in radians) e) The spherical vector illustration of the time evolution of a spin at fixed $x=0$ displaying a nutation f) The snapshots of the spin chain at $t_0=0$ and $t_1=2$ showing a propagating nutation wave (blue arrow shows the wave propagation direction along $x$). The spherical coordinate frame for  $\mathbf{M}$  uses the polar angle $\theta$ and azimuthal angle $\phi$ as per standard conventions. One unit of distance (time) in the figures corresponds to $a=2\AA$($\tau=10^{-14}s$).}
 \end{figure}

In order to verify that such propagating wave solutions exist, beyond the linear approximation, we solved eq.(\ref{ILLGspatial0}) numerically. To isolate the nutation wave and suppress the spin wave, we set the DC magnetic field to zero $\mathbf{H}_{\mathrm{DC}}=0$  while applying an electromagnetic wave, described by  $\mathbf{H}_{\mathrm{AC}}$ with a frequency that should match that of the nutation wave (resonance frequency). The electromagnetic wave driving field serves to excite the nutation motion and drive it  across the spin chain. Fig. 1. obtained from solving the equation around a static background as shown in Fig. 2. clearly shows the occurrence of the nutation wave in terms of the spherical angles $\theta(x,t)$ and $\phi(x,t)$ (a spin wave would only display a periodic pattern in $\phi(x,t)$). The existence of the nutation wave is made possible by the exchange field that ``mediates" the propagation of this wave.
\begin{figure}
 \includegraphics[angle=0,origin=c, scale=0.125]{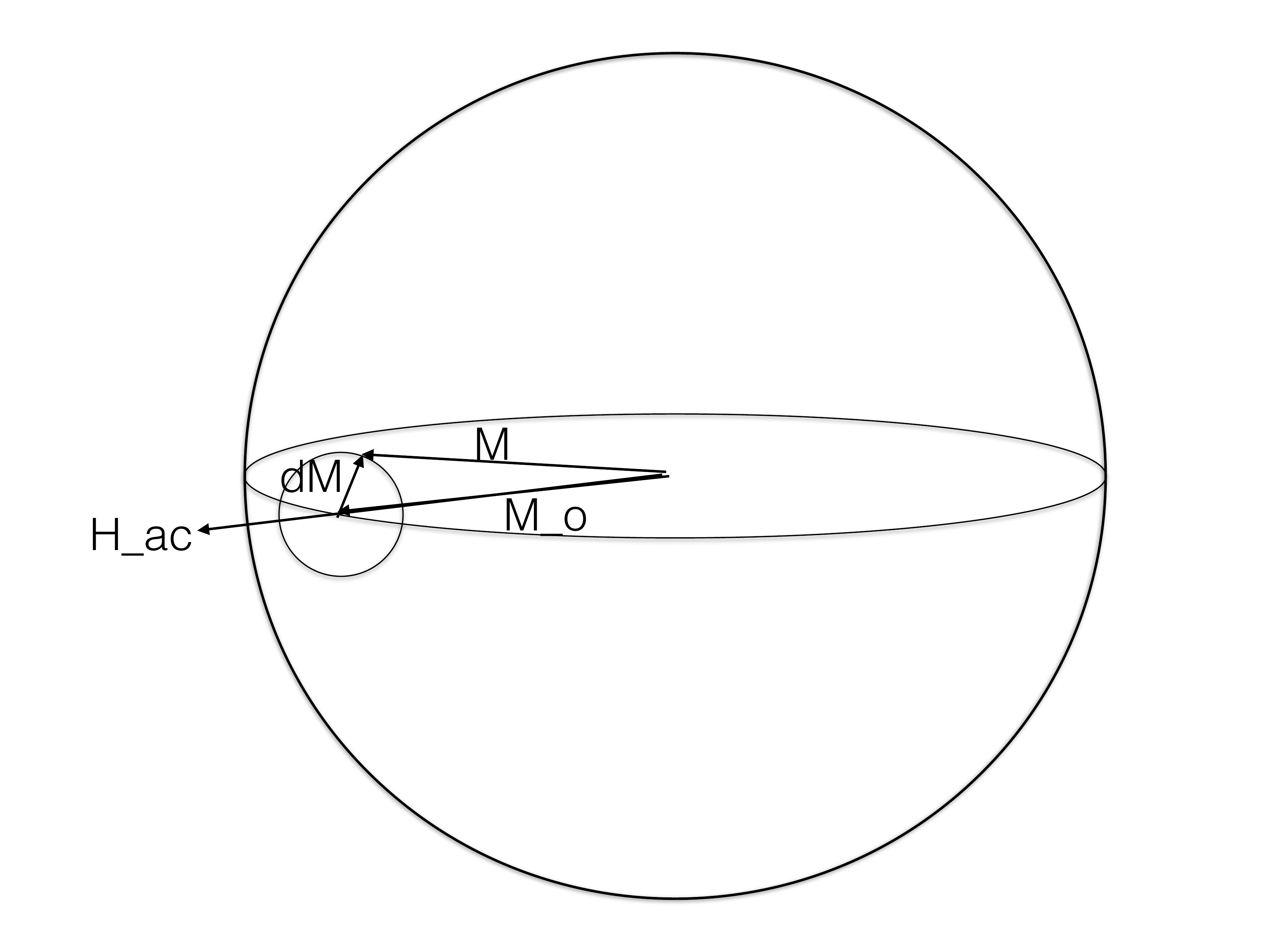}
 \label{fig:NutatonProfile}
 \caption{
 The profile of magnetization vector with nutation around the equatorial plane.}
 \end{figure}

We now derive the dispersion relation of the nutation wave as follows. We consider a situation where the equilibrium magnetization is oriented around the equatorial plane of the sphere where the magnetization vector lives, as illustrated in Fig. 2. In most cases of interest, the nutation effect is normally much smaller in magnitude than principal magnetization that undergoes the precession. It is therefore legitimate to linearize the equation of motion eq.(\ref{ILLGspatial0}) by using eq.(\ref{expanding}) which describes only nutation with no precession when $\mathbf{H}_{\mathrm{DC}}=0$. 

We obtain the dispersion relation for the collective modes by expanding the magnetization vector $\mathbf{M}$ in eq.(\ref{ILLGspatial0}) using eq.(\ref{expanding}) but renaming $\tilde{\mathbf{M}}$ with $\delta\mathbf{M}$, retaining terms up to linear order in $\tilde{M}$, and then taking the Fourier transform, giving 
\begin{equation}
\label{dispersionrel}
\begin{array}{l}
\mathrm{det}\left[\mathrm{i}\omega\delta_{IJ}-( \mathrm{i}\gamma\eta\omega+\gamma\eta\tau\omega^2-\gamma\eta\tau v^2_\mathrm{iner}k^2  )\varepsilon_{IJK}M_{0,K}\right]=0
\end{array}
\end{equation}
where "det" represents the determinant of the $3\times 3$ matrix in its argument and $I,J,K=x,y,z$ label the components of the vector $\mathbf{M}$, $\sum_{K}M^2_{0,K}=M^2_s$ while $\varepsilon_{IJK}$ is the totally antisymmetric tensor. This equation is further subject to the constraint $|\mathbf{M}|^2=M^2_s$ which gives
\begin{equation}
    \mathbf{M}_0\cdot \delta\mathbf{M}=0
\end{equation}
to linear order in $ \delta\mathbf{M}$ which simply means that the fluctuation of the magnetization $\delta\mathbf{M}$ is normal to the the static background $\mathbf{M}_0$, as illustrated in Fig. 2. The constraint reduces the effective matrix into $2\times 2$ matrix and results in the following polynomial equation for $\omega$
\begin{equation}\label{dispersionrelationequation}
    \omega^2=\left(\omega_k-\alpha\omega(i+\omega\tau)\right)^2
\end{equation}
where
\begin{equation}
    \omega_k=\gamma J M_s k^2
\end{equation}
Equation (\ref{dispersionrelationequation}) is a fourth-order polynomial equation with four solutions for $\omega$ in terms of $k$. They consist of two pairs of solutions; the two pairs differ only by opposite directions of propagation. We will only give the equations for the right-moving solutions, whose (complex-valued) dispersion relations are given by  
\begin{equation}\label{complexdispersions}
    \omega_{\pm}(k)=\frac{-(\pm 1+i \alpha)+\sqrt{4\alpha\tau\omega_k+(\pm 1+i\alpha)^2}}{2\alpha\tau}
\end{equation}
We will first focus on the real part of the complex energy dispersions eqn. (\ref{complexdispersions}). 

The first solution $\omega_{+}(k)$ turns out to give a non-relativistic (quadratic) dispersion relation associated with a precession mode
\begin{equation}\label{spinwavedispersionILLGspatial}
\epsilon_{\mathrm{sw}}(k)=\mathrm{Re}[\omega_{+}(k)]=\gamma J M_s k^2
\end{equation}
at small wave vector $k$, which corresponds to the standard form of a spin wave dispersion. The important point is that this dispersion is gapless; i.e.$\epsilon_{\mathrm{sw}}(k=0)=0$. 

The second solution $\omega_{-}(k)$ turns out to give a massive dispersion relation. Its real part is given by 
\begin{equation}\label{realdispersionnutationwave}
\epsilon_{\mathrm{nw}}(k)=\mathrm{Re}[\omega_{-}(k)]\simeq\frac{1}{2\alpha\tau}+\sqrt{v^2_{\mathrm{iner}}k^2+\frac{1}{4\alpha^2\tau^2}}
\end{equation}
where 
\begin{equation}
\alpha=\gamma M_s\eta
\end{equation}
with a gap 
\begin{equation}\label{nutatongap}
m=\frac{1}{\alpha\tau}
\end{equation}
We interpret this second solution as a propagation of the nutation; that is, the nutation wave that turns out to have a massive relativistic dispersion, having single-particle excitations (the "nutatons") with mass $m$. The positive definiteness of the gap guarantees the stability of the nutation wave in the sense that it well separates from low-energy excitation.

Writing the full complex wave dispersion relations as
\begin{equation}
    \omega_{\pm}(k)=\omega^R_{\pm}(k)-i\omega^I_{\pm}(k)
\end{equation}
where $\omega^R_{+(-)}(k)=\epsilon_{\mathrm{sw}(\mathrm{nw})}(k)$, the imaginary part $\omega^I_{+}(k)$ for spin wave is found to be
\begin{equation}
    \omega^I_{+}(k)=\alpha\omega_k=K k^2
\end{equation}
where $K=\alpha \gamma J M_s$, which increases with the wave vector $k$, which implies that a zero wave vector spin wave never decays while it decays faster for larger wave vectors or shorter wave lengths. On the other hand, for the nutation wave we obtain
\begin{equation}\label{imaginarypartNW}
    \omega^I_{-}(k)=\frac{1}{\tau}-\alpha\omega_k=\alpha m-Kk^2
\end{equation}
which implies that the zero wave vector nutation wave actually decays faster and as the wave vector increases (or wave length decreases), the nutation wave decays slower, or in other words, becomes even more robust (slowly-decaying), which is a rather surprising result.  

More precisely, the robustness of the nutation wave can be described by the ratio between the imaginary and real parts
\begin{equation}
    r(k)=\frac{\omega^I_{-}(k)}{\omega^R_{-}(k)}=\frac{\alpha m-K k^2}{\frac{m}{2}+\sqrt{{v}^2_{\mathrm{iner}}k^2+{\left(\frac{m}{2}\right)}^2}}
\end{equation}
Physically, $1/r(k)$ gives the number of periods that a nutation wave with wave vector $k$ covers as it propagates and decays with time. Defining $k=(m/(2v_{\mathrm{iner}}))\kappa=k_0 \kappa$, where $\kappa$ is a dimensionless wave vector parameter and 
\begin{equation}
k_0=\frac{m}{2v_{\mathrm{iner}}}
\end{equation}
is the crossover wave vector from quadratic to linear dispersion, we have
\begin{equation}
    r(\kappa)=2\alpha\frac{1-\frac{\kappa^2}{4}}{1+\sqrt{\kappa^2+1}}
\end{equation}
Since $\alpha\lesssim 0.1$ in real materials and the dispersion is practically linear once $\kappa\gtrsim 1$ as described by the group velocity 
\begin{equation}
    v_g(k)=\frac{\partial \omega^R_{-}(k)}{\partial k}=v_{\mathrm{iner}}\frac{\kappa}{\sqrt{\kappa^2+1}}
\end{equation}
that already gives $v_g\simeq v_{\mathrm{iner}}$ for $\kappa\gtrsim 1$, we conclude that $r\ll 1$ indeed, indicating that we have a high speed nutation wave that will be robust against decay within time scale for observation $\tau_d=2\pi/\omega^I_{-}(k_0)\gg \tau, 2\pi/\omega^R_{-}(k_0)$ and over a length scale $l_d\sim v_{\mathrm{iner}}\tau_d$. Indeed for realistic materials parameters that we will exemplify at the end of this section, we find that 
\begin{equation}
    l_d=\frac{8\pi v_{\mathrm{iner}}\tau}{3}\gg a
\end{equation}
where $a$ is the lattice constant of typical spin chain materials. Equation (\ref{imaginarypartNW}) also imposes that the robustness of the nutation wave requires $|k|\leq 2k_0$, thus giving 
\begin{equation}
    k_{\mathrm{UV}}=2k_0
\end{equation}
as a natural ultraviolet cutoff of our theory, which also defines the regime of validity of our continuum theory defined by equations (\ref{ILLG})-(\ref{ILLGspatial0}).

The nutation wave dispersion relation~eq.(\ref{realdispersionnutationwave}) describes  a  relativistic massive particle, which thus defines a completely different excitation  from that of a spin wave, which is gapless and non-relativistic. A spin wave can thereby transport arbitrarily small energy across the spin chain. On the other hand, a nutaton needs a finite energy to move even at infinitely small wave vector as illustrated in Fig. 3. 

\begin{figure}[thp]
 \includegraphics[angle=0,origin=c, scale=0.50]{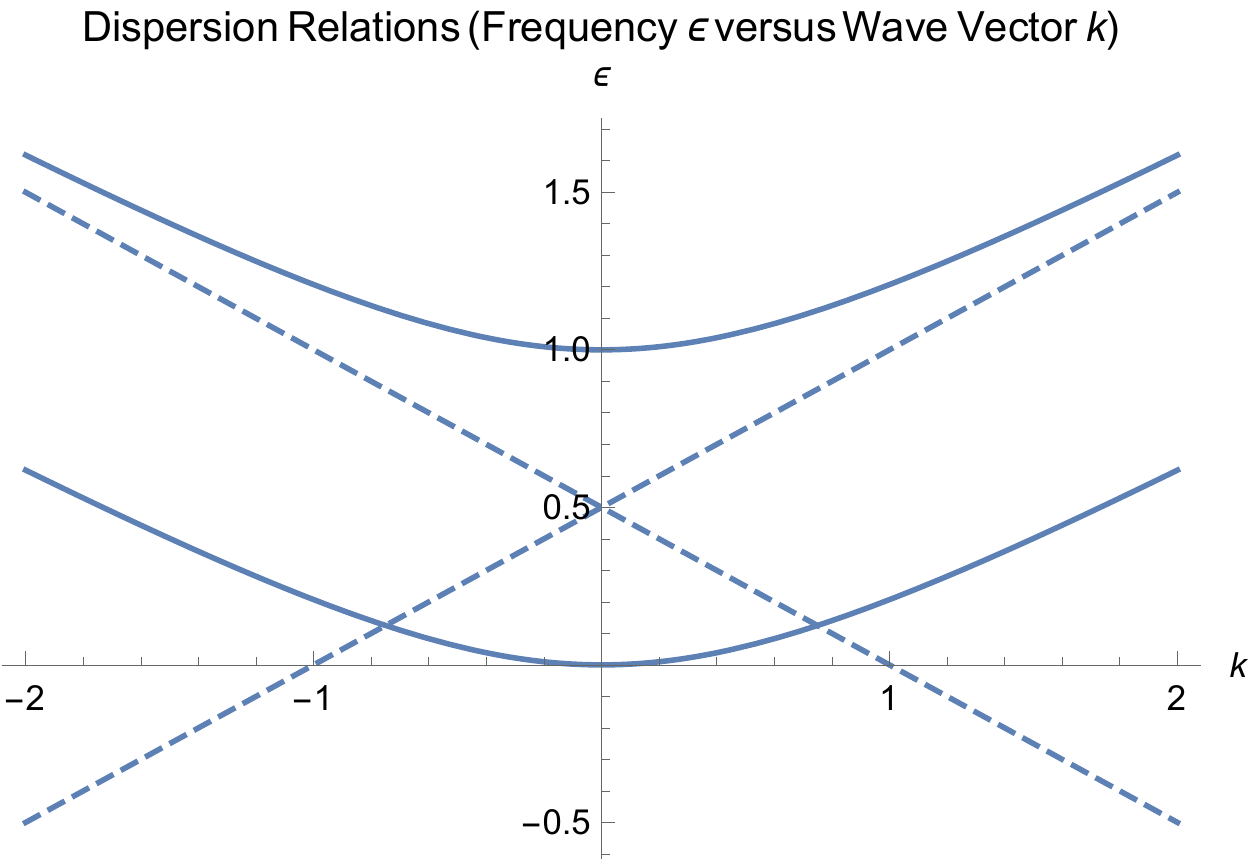}
 \label{fig:NutatonEnergyDispersion}
 \caption{
 An illustrative profile of the dispersion relations of the nutation wave $\epsilon_{\mathrm{nw}}(k)$ (upper curve) and of spin wave $\epsilon_{\mathrm{sw}}({k})$ (lower curve) where the units are $m=10^{14}$ Hz for the frequency and $k_0=0.714{\AA}^{-1}$ for the wave vector $k$. The dashed straight lines are guides to the eye on the linear-in-$k$ part of the nutation wave dispersion at large $k$.}
 \end{figure}

We now compare the characteristic frequencies of the two waves. At the crossover wave vector $k_0$, the nutation wave frequency is found to be
\begin{equation}\label{nutationcrossoverenergy}
    \epsilon_{\mathrm{nw}}(k_0)=\frac{1+\sqrt{2}}{2}m
\end{equation} 
On the other hand, we find that for the spin wave
\begin{equation}
    \epsilon_{\mathrm{sw}}(k_0)=\frac{1}{4}m
\end{equation}
The last two equations indicate that the nutation wave \textit{always} has higher frequency than spin wave when the former enters the linear part of its dispersion. 

Our results can be directly verified in one-dimensional or quasi-one-dimensional magnetic materials such as $\mathrm{CuCl_2-TMSO}$, $\mathrm{C_4H_8SO(TMSO)}$, $\mathrm{C_2H_6SO(DMSO)}$ and $\mathrm{(C_6H_{11}NH_3)CuCl_3}$ \cite{HeisenbergFerromagnetSpinchain1}\cite{HeisenbergFerromagnetSpinchain2}\cite{HeisenbergFerromagnetSpinchain3}. Considering $\mathrm{C_4H_8SO(TMSO)}$ and $\mathrm{C_2H_6SO(DMSO)}$, the best fit gave exchange constants $J_{\mathrm{exch}}/k_B=39$K and $45$K (where $k_B$ is the Boltzmann constant) and lattice spacing $a\simeq 2 \AA$  \cite{HeisenbergFerromagnetSpinchain2}$^{**}$.As an illustration, using $J=J_{\mathrm{exch}}/(M^2_s a)$, $\gamma=1.75\times 10^{11}$ sA/kg, $\tau=10^{-14}s$, $M_s=10^6$ A/m and $\alpha=\gamma \eta M_s=0.1$ \cite{APL}, we obtain the nutation wave resonance frequency at $k=k_0$ to be $\epsilon_{\mathrm{nw}}(k_0)\simeq 1.207\times 10^{15}$ Hz and characteristic velocity $v_{\mathrm{iner}}=\sqrt{\gamma JM_s/(\tau\alpha)}\simeq 22\times 10^3$ m/s which is nearly one order of magnitude larger than the characteristic group velocity of a spin wave $v_{\mathrm{sw}}=\partial \omega_{\mathrm{sw}}(k)/\partial k_{|k=k_{\mathrm{sw}}}\simeq 3\times 10^{3}$ m/s corresponding to the typical value known in literature (e.g. \cite{NatureNano}\cite{NatComms}), with $k_{\mathrm{sw}}\simeq 0.14 k_0\ll \pi/a$.This estimate thus validates our conclusion regarding the primacy of the nutation wave in terms of its characteristic speed and frequency as well as robustness. In fact, using the realistic materials parameters above, we do indeed find $r\ll 1, l_d\gg a$ and $k_0\ll \pi/a$ (with larger $\tau$ and $J$) that justifies our results. As it turns out, larger values for $J$ and $\tau$ (e.g. $\tau=10^{-12}$s) makes the conditions $l_d\gg a$ and $k_0\ll \pi/a$ even better satisfied \textit{simultaneously}, indicating the self-consistency of our theory and its results. Typical values of $\alpha$ are smaller than what we used above \cite{Neeraj}, which would still make the nutation wave characteristic velocity $v_{\mathrm{iner}}$ larger than that of the spin wave $v_{\mathrm{sw}}$. Using $\alpha=0.01, \tau=10^{-12}$s for example, we obtain $\epsilon_{\mathrm{nw}}(k_0)\simeq 1.207\times 10^{14}$, $v_{\mathrm{iner}}\simeq 7\times 10^3$m/s as illustrated in Fig. 3., and $l_d=586\AA,r=0.006,k_0=0.714{\AA}^{-1}$, satisfying all the above constraints. 

The apparent separation of (inverse) length scale ($k_{\mathrm{sw}}$ for spin wave and $k_0$ for nutation wave) originates from the corresponding separation in the time scale ($t\ll \tau$ for nutation wave and $t\gg \tau$ for spin wave). Furthermore, our robustness analysis suggests that the spin wave is less robust (faster-decaying) at larger wave vectors when the nutation wave is more robust (slowly-decaying). The spin wave is thus limited to have long wave lengths while nutation can potentially have much shorter wave lengths. In other words, in Fig. 3., the spin wave dispersion is actually constrained to cover only very small wave vectors $|k|\leq k_{\mathrm{sw}}$ around $k=0$ while nutation wave covers much larger interval of wave vectors $|k|\lesssim 2k_0$, corresponding to shorter wave lengths. 

 \begin{figure}[thp]
 \includegraphics[angle=0,origin=c, scale=0.15]{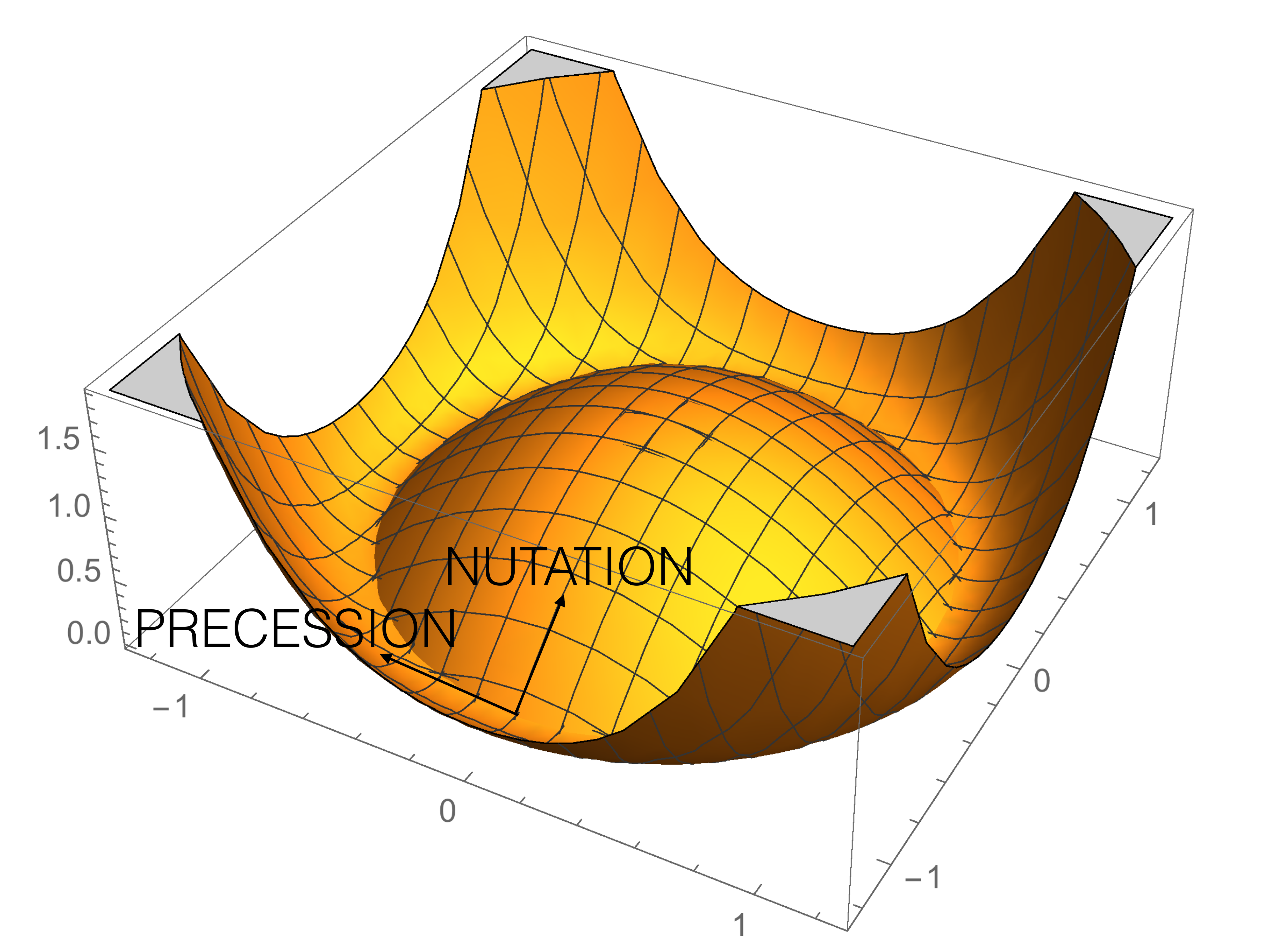}
 \label{fig:HiggsModeNutaton}
 \caption{
A $\Phi^4$ potential superimposed on the sphere in Fig.2. describing the precession along azimuthal and nutation away from equator. The horizontal (vertical) axes (axis) represent the order parameters (energy).}
 \end{figure}

We argue below that the appearance of the gap is a realization of the Brout-Englert-Higgs mechanism \cite{Higgs1}'\cite{Higgs2} where the nutaton acts as a Higgs mode, described by $M_{\perp}$, analogous to the Higgs amplitude modes in superconductor \cite{Sherman} or quantum magnetism \cite{Wessel}'\cite{Sachdev}'\cite{PekkerVarma}. It will be demonstrated that the long-distance (small $k$) dynamics of the ferromagnets can be described by the standard Lagrangian describing the Abelian Higgs model \cite{PeskinSchroederQFT}
\begin{equation}\label{GaugeFieldTheory}
\mathcal{L}=-\frac{1}{4}F^2_{\mu\nu}+|D_{\mu}\Phi|^2-V(\Phi)
\end{equation}
with field tensor $F_{\mu\nu}=\partial_{\mu}A_{\nu}-\partial_{\nu}A_{\mu}$ and covariant derivative $D_{\mu}=\partial_{\mu}+ieA_{\mu}$. Considering static background magnetization along $x$ as in earlier part, the complex scalar field $\Phi$ is given by
\begin{equation}\label{scalarfield}
\Phi=M_x+iM_{\perp}=M_se^{i\chi}
\end{equation}
\begin{equation}\label{definitions}
M_{\perp}=\sqrt{M^2_y+M^2_z},\chi=\tan^{-1}\frac{M_{\perp}}{M_x}
\end{equation}
The Abelian gauge field here is an emergent (i.e. not the physical electromagnetic) gauge field given by $A_{\mu}=-\frac{1}{e}\partial_{\mu}\chi$ and $e$ is the corresponding effective gauge charge. This gauge field thus represents the spatial and temporal variations (i.e. gauge fluctuations) of the spin orientation \cite{MakhfudzPRB2014}'\cite{MakhfudzJPCM}; the phase fluctuations of $\Phi$. The Lagrangian is both Lorentz and gauge invariant under the gauge transformation
\begin{equation}
\Phi\rightarrow e^{i\xi}\Phi, A_{\mu}\rightarrow A_{\mu}-\frac{1}{e}\partial_{\mu}\xi
\end{equation}
which corresponds to the rotation of the magnetization vector $\mathbf{M}$ on a sphere of radius $M_s$ \cite{MakhfudzJPCM} that brings $\mathbf{M}=(M_x,\mathbf{M}_{\perp})$ to $\mathbf{M}'=(M'_x,\mathbf{M}'_{\perp})$ and gives
\begin{equation}
\xi=\tan^{-1}\frac{M'_{\perp}M_x-M_{\perp}M'_x}{M_xM'_x+M_{\perp}M'_{\perp}}
\end{equation}
where $\mathbf{M}_{\perp}=(M_y,M_z), \mathbf{M}'_{\perp}=(M'_y,M'_z)$ and $|\mathbf{M}|=|\mathbf{M}'|=M_s$. Note that for our one-dimensional spin system (spin chain), only $\mu,\nu=0,1\equiv t,x$ and $A_t,A_x$ are present and as a result, the $F_{\mu\nu}$ only gives $E_x=-\partial_xA_t-\partial_tA_x$ while all other field tensor components vanish.

The potential $V(\Phi)$ is that of the  standard $\Phi^4$ theory illustrated in Fig. 4. 
\begin{equation}\label{Potential}
V(\Phi)=-\mu^2 |\Phi|^2+\frac{\lambda}{2}|\Phi|^4
\end{equation}
Now consider the  symmetry breaking due to  a vacuum expectation value of eq.(\ref{VEV}) given by $\Phi_0=iM_s$ such that
\begin{equation}\label{VEV}
|\Phi_0|=|\langle \Phi\rangle|=\sqrt{\frac{\mu^2}{\lambda}}
\end{equation} 
This describes magnetization, that is  uniformly oriented on the $xy$ (equatorial) plane in the vector $\mathbf{M}$ space in Fig. 2.  

We add fluctuations around this symmetry-breaking ground state
\begin{equation}\label{perturbedfield}
\Phi(x)=\Phi_0+\frac{1}{\sqrt{2}}(\Phi_1(x)-i\Phi_2(x))
\end{equation}
where we note by comparing with eq.(\ref{scalarfield}) that $\Phi_1(x)/\sqrt{2}\equiv \delta M_x$ while $-\Phi_2(x)/\sqrt{2}\equiv\delta M_{\perp}$. Now, consider a functional subspace for which
\begin{equation}
\Phi_2(x)=\sqrt{2}M_s\left(1-\sqrt{1-\frac{\Phi^2_1(x)}{2M^2_s}}\right)
\end{equation}
This configuration satisfies the constraint
\begin{equation}\label{constraint}
|\Phi(x)|^2=M^2_s
\end{equation}
representing a functional subspace for which the fluctuations eq.(\ref{perturbedfield}) preserve the magnitude $|\mathbf{M}|=M_s$, corresponding to the sphere in Fig. 2. Therefore, the fluctuation of eq.(\ref{perturbedfield}) represents an amplitude mode \cite{Mannheim} in terms of $M_{\perp}$ rather than $|\mathbf{M}|=M_s$ as the latter is fixed. Substituting eq.(\ref{perturbedfield}) into eq.(\ref{Potential}), we obtain
\begin{equation}\label{EffectivePotential}
V(\Phi_1,\Phi_2)=-\frac{\mu^4}{2\lambda}+\mu^2\Phi^2_2+\mathcal{O}(\Phi^3_1,\Phi^3_2)
\end{equation}
Clearly, the $\Phi_2$ field that changes the $M_{\perp}$ acquires a mass $m^2_2=2\mu^2$, which according to eqs.(\ref{GaugeFieldTheory}) and (\ref{EffectivePotential}) gives a dispersion of the form $\sim k^2+\mu^2$ at small $k$, corresponding to the nutation wave acting as a Higgs mode. The $\Phi_1$ remains massless with dispersion $\sim k^2$ corresponding to the spin wave as a Goldstone mode, in complete agreement with the analysis of the dispersion relations. The gauge field $A_{\mu}=-(\partial_t\chi,\partial_x\chi)/e$ acquires a mass term $\delta\mathcal{L}=\frac{1}{2}m^2_AA^2_{\mu}$ where $m^2_A=2e^2\frac{\mu^2}{\lambda}$. The phase fluctuating field thus acts as the gauge vector boson that gains mass and acquires one degree of freedom via this mechanism.

The nutation wave would appear as a peak, along with a gap, in the structure factor $S(\omega,k)$, as measured in inelastic neutron scattering experiments where the peak is given by $\omega=\epsilon_{\mathrm{nw}}(k)$. From a practical point of view, the nutation wave may have significant consequences in the field of spintronics. In particular, despite being smaller in amplitude, compared to precession spin wave \cite{JAP}, the nutation wave has much higher characteristic frequencies than the latter. Furthermore, the robustness of the nutation wave at short wavelengths is perfect for spintronic applications, such as in realizing much faster magnetic switching processes~\cite{nussle2019dynamic}. 

%\newline
The data that supports the findings of this study are available within the article [and its supplementary materials].

Supplementary Material

See the supplementary materials for the details of the analytical derivations, numerical calculation, and their analyses.

I. M. thanks GREMAN for a work travel financial support useful to this research. The authors thank Prof. J-C. Soret for critical reading of the manuscript and important feedback.

\newpage

\begin{widetext}

\centering
\textbf{Supplementary Materials: Nutation Wave as a Platform for Ultrafast Spin Dynamics in Ferromagnets}\\% Force line breaks with \\
\centering
Imam Makhfudz$^{\psi}$ Enrick Olive$^{\psi}$ and Stam Nicolis$^\phi$\\
%\affiliation{%
\centering
$^{\psi}$GREMAN, UMR 7347, Universit\'{e} de Tours-CNRS, INSA Centre Val de Loire, Parc de Grandmont, 37200 Tours, France\\ 
$^{\phi}$Institut Denis Poisson, Universit\'{e} de Tours, Universit\'{e} d'Orl\'{e}ans, CNRS (UMR7013), Parc de Grandmont, F-37200, Tours, France
 %}%
\date{\today}% It is always \today, today,
\bigskip

In these Supplementary Materials, we first give the derivation of the linearized form of the inertial Landau-Lifshitz-Gilbert (ILLG) equation. Then the derivation of the dispersion relations of the nutation and spin waves. Next, the details of the numerical calculation. Finally, additional sections discussing briefly the strength of the nutation wave, comparison between dynamical and topological inertia, and an additional note on the nutation wave dispersion relation.

\end{widetext}

\section{Linearized Landau-Lifshitz-Gilbert Equation}
Anticipating harmonic solutions with well-defined velocity, we rewrite Eq.(5) in the main text in space-time translational invariant form by expressing the total time derivative in terms of partial derivatives
\begin{equation}\label{totalderivatives1}
\frac{d\tilde{\mathbf{M}}(x,t)}{dt}=\frac{\partial \tilde{\mathbf{M}}}{\partial t} +\frac{d x}{d t}\frac{\partial \tilde{\mathbf{M}}}{\partial x}=\frac{\partial \tilde{\mathbf{M}}}{\partial t} \pm v_g\frac{\partial \tilde{\mathbf{M}}}{\partial x}=D_g \tilde{\mathbf{M}} 
\end{equation}
\begin{equation}\label{totalderivatives2}
\frac{d^2\tilde{\mathbf{M}}(x,t)}{dt^2}=\frac{\partial^2\tilde{\mathbf{M}}}{\partial t^2}\pm 2v_g\frac{\partial^2\tilde{\mathbf{M}}}{\partial x\partial t}+v^2_g\frac{\partial^2\tilde{\mathbf{M}}}{\partial x^2}=D^2_g\tilde{\mathbf{M}}
\end{equation}
where $v_g$ is an averaged group velocity. 

In equation Eq.(8) in the main text, the relations between the transformed variables and original variables are derived as follows. The last two terms in Eq.(5) in the main text upon linearization becomes 
\begin{equation}\label{relativisticpart}
\frac{\partial^2\tilde{\mathbf{M}}}{\partial t^2}\pm 2v_g\frac{\partial^2\tilde{\mathbf{M}}}{\partial x\partial t}+(v^2_g-v^2_{\mathrm{iner}})\frac{\partial^2\tilde{\mathbf{M}}}{\partial x^2}
\end{equation}
The resulting linearized equation constitutes a nontrivial partial differential equation for the dynamical vector field $\tilde{\mathbf{M}}(x,t)$ which contains both non-relativistic and relativistic parts. To expose the relativistic character, we perform a Lorentz boost transformation along $x$
\begin{equation}
t=\frac{1}{\sqrt{1-\frac{v^2}{c^2}}}\left(t'+ \frac{v}{c^2} x'\right), x=\frac{1}{\sqrt{1-\frac{v^2}{c^2}}}\left(x'+ v t'\right)
\end{equation}
where the appropriate $v$ and ``speed of light" $c$ are to be determined. We factorize Eq. (\ref{relativisticpart}):  
\[
\frac{\partial^2\tilde{\mathbf{M}}}{\partial t^2}\pm 2v_g\frac{\partial^2\tilde{\mathbf{M}}}{\partial x\partial t}+(v^2_g-v^2_{\mathrm{iner}})\frac{\partial^2\tilde{\mathbf{M}}}{\partial x^2}
\]
\begin{equation}\label{equality}
=\left(\frac{\partial }{\partial t'}-\tilde{v}_{s=+}\frac{\partial }{\partial x'}\right)\left(\frac{\partial }{\partial t'}-\tilde{v}_{s=-}\frac{\partial }{\partial x'}\right)\tilde{\mathbf{M}}=D^2_s\tilde{\mathbf{M}}
\end{equation}
where the velocities $\tilde{v}_{s=\pm}$ given by 
\begin{equation}
\tilde{v}_{s=\pm}=s v_{\mathrm{iner}}\pm v_g
\end{equation}
are those of the wave equation defined by setting both sides of Eq.(\ref{equality}) to zero. We perform Fourier transform on both sides of the equation 
\begin{equation}
\tilde{\mathbf{M}}(x,t)=\int \frac{d\omega dk}{(2\pi)^2} \tilde{\mathbf{M}}(k,\omega)e^{i(kx-\omega t)}
\end{equation}
and make use of the Lorentz boost transformation to arrive at a relation between the Fourier transform of the two sides that gives
\begin{widetext}
\begin{equation}
-\left(k \gamma v - \omega \gamma\right)^2 + \left(\tilde{v}_{s=+} + \tilde{v}_{s=-}\right) \left(k \gamma v - \omega \gamma\right) \left(k \gamma - \frac{\omega \gamma v}{c^2}\right) - \tilde{v}_{s=+} \tilde{v}_{s=-} \left(k \gamma - \frac{\omega \gamma v}{c^2}\right)^2 + \omega^2 - 
 2 v_g \omega k + v^2_g k^2 - v^2_{\mathrm{iner}} k^2=0
\end{equation}
\end{widetext}
and matching  powers of $\omega$, we obtain
\begin{equation}
v = 0, v = -\frac{2 c^2 v_g}{c^2 + v^2_g - v^2_{\mathrm{iner}}},v = \frac{-c^2 - v^2_g + v^2_{\mathrm{iner}}}{2 v_g}
\end{equation}
Equating the two nontrivial solutions for $v$ gives 
\begin{equation}
c=v_g\pm v_{\mathrm{iner}}
\end{equation}
Eventually, as we will see later, we can define $v_g$ to be an averaged group velocity of the wave solution of Eq. (5) in the main text while $v$ itself is the group velocity of each mode: $v=v_g(k)$ of the excitation of our interest. Since we are mainly interested in the nutation wave, which turns out to have a massive relativistic particle spectrum, we use $v=v_g(k)$ for the nutation wave (which we will compute in the next section) which means that we aboard a frame of reference attached to a Fourier mode of the nutation wave with wave vector $k$. We will show that $v_g= K v_{\mathrm{iner}}$ where $0<K\leq 1$ upon the application of electromagnetic wave and as a result, $c=(K\pm 1)v_{\mathrm{iner}}$. Since the maximum group of the nutaton is $v^{\mathrm{max}}_g=v_{\mathrm{iner}}$, we consider $c\leq v_{\mathrm{iner}}$. When $|K\pm 1|=1/2$, we have $|\tilde{v}_g|,|\tilde{v}_{s=\pm}|=v_{\mathrm{iner}}/2$ which makes Eq.(8) in the main text manifest not only space-time translational invariance but also the Lorentz invariance.

The scaled variables in the linearized equation are given by
\begin{equation}
\tau'=\frac{\tau}{\sqrt{1-\frac{v^2}{c^2}}}
\end{equation}
\begin{equation}
\tilde{\mathbf{H}}=\frac{\mathbf{H}}{\sqrt{1-\frac{v^2}{c^2}}}
\end{equation}
while
\begin{equation}\label{wavevelocityS}
v_{\mathrm{iner}}=\sqrt{\frac{J}{\eta\tau}}
\end{equation}
The first derivative terms on both sides of the Eq.(5) in the main text become
\begin{equation}
\frac{\partial \tilde{\mathbf{M}}}{\partial t} \pm v_g\frac{\partial \tilde{\mathbf{M}}}{\partial x} \rightarrow \frac{\partial \tilde{\mathbf{M}}}{\partial t'} \pm \tilde{v}_g\frac{\partial \tilde{\mathbf{M}}}{\partial x'}
\end{equation}
where $\tilde{v}_g=v_g$. One should note Eq.(8) in the main text represents a form of non-trivial driven vector wave equation with non-relativistic and relativistic parts, representing the spin wave and nutation wave respectively. In this case, we do not have one constant of motion defining the velocity of the wave solution like that in electrodynamics. Only at one special point when $|\tilde{v}_g|=|\tilde{v}_s|$ do we have  manifest Lorentz invariance. The average group velocity $\tilde{v}_g$ also represents a statistically averaged velocity of a group of modes rather than that of any one  mode. Despite the absence of exact Lorentz invariance corresponding to one unique wave solution, Eq. (8) still nevertheless admits wave-type of solutions, one of which corresponds to a relativistic wave solution called the nutation wave.

\section{Derivation of the Dispersion Relation of the Nutation Wave}
We now derive the dispersion relation of the nutation wave by linearizing the equation of motion Eq.(5) as follows. The magnetization vector $\mathbf{M}$ is oriented along a direction in the sphere in which it lives, as illustrated in Fig. 5. where we display only a single spin as a representative figure. This vector  $\mathbf{M}$ may precess around a static magnetic field $\mathbf{H}_{\mathrm{DC}}$ and also exhibit a nutation on top of it generated by an alternating magnetic field $\mathbf{H}_{\mathrm{AC}}$. 
\begin{figure}
 \includegraphics[angle=0,origin=c, scale=0.20]{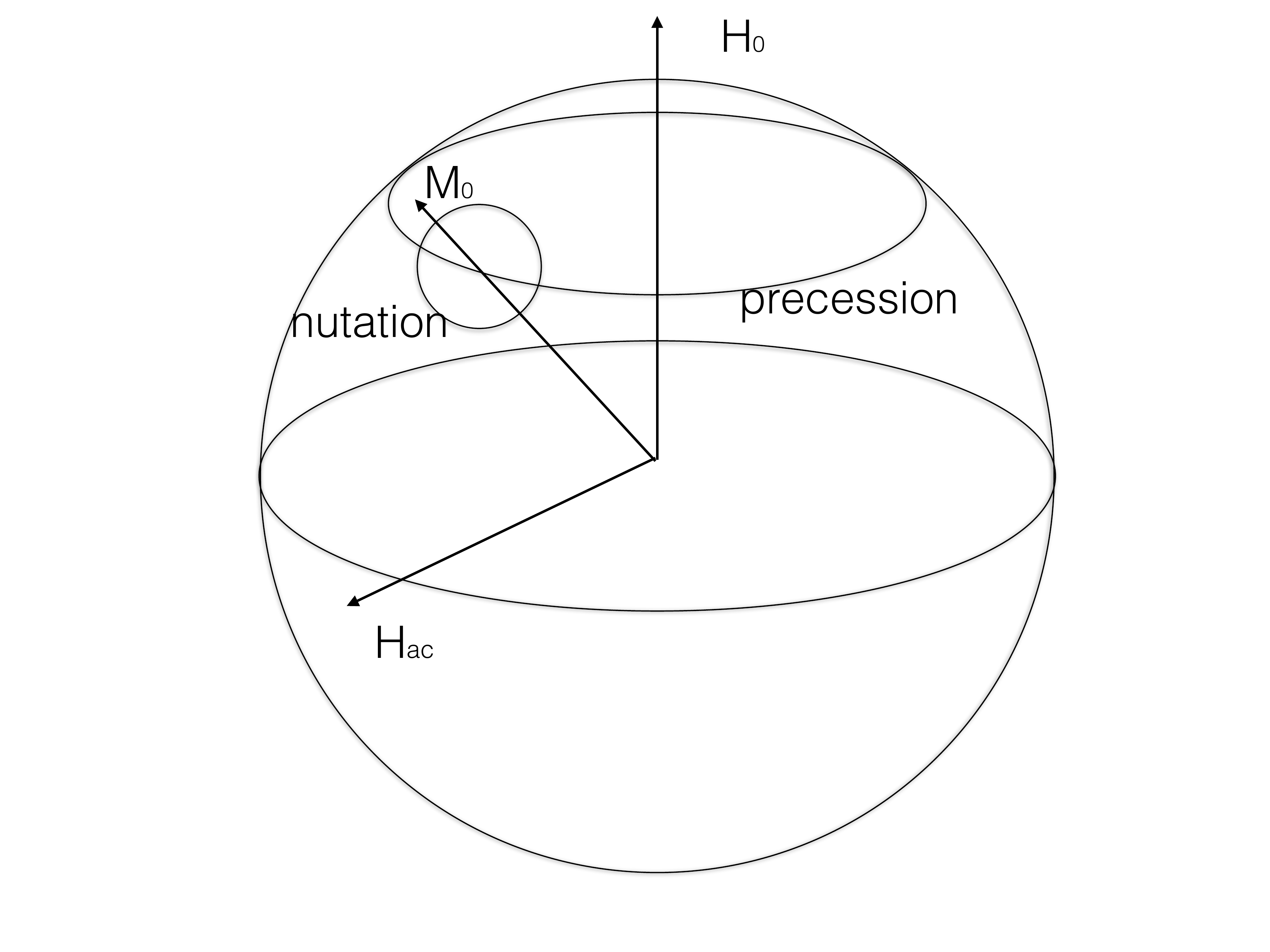}
 \label{fig:NutatonProfile}
 \caption{
 The profile of a magnetization vector defined on a sphere, undergoing precession and nutation at the same time.}
 \end{figure}
We will apply a static (DC) magnetic field along $z$; $\mathbf{H}_{\mathrm{DC}}=H_0\hat{\mathbf{z}}$ and an electromagnetic wave which carries an AC magnetic field $\mathbf{H}_{\mathrm{AC}}$ polarized along $x$; $\mathbf{H}_{\mathrm{AC}}=H_{\mathrm{ac}}(x,t)\hat{\mathbf{x}}$. Both fields will be taken to zero at the end of the calculations. The AC driving field serves to excite the nutation motion as well as drives it across the spin chain. The simultaneous application of static and alternating magnetic fields will generate both precessional and nutational modes \cite{APLs}\cite{JAPs}. In most cases of interest, the nutation effect is normally much smaller in magnitude than principal magnetization that undergoes the precession. It is therefore legitimate to perform linearization on the equation of motion Eq.(5) by writing
\begin{equation}\label{linearanalysis}
\mathbf{M}(x,t)=\mathbf{M}_0+\delta\mathbf{M}(x,t)
\end{equation}
where $\mathbf{M}_0$ is the static background magnetization and $\delta\mathbf{M}(x,t)$ is the fluctuation around it. We will choose $\mathbf{M}_0$ to be along a given direction for concreteness but the final result does not depend on the choice because we consider an isotropic ferromagnet within exchange approximation.

Performing the Fourier transform on Eq.(5) upon linearization, we obtain the following matrix equation
\begin{widetext}
\[
\begin{pmatrix}
 0 & -|\gamma_0|H_0 & 0 \\
  |\gamma_0|H_0 & 0 & -(\omega_k-\alpha\omega(i+\omega \tau)) \\
  0  & (\omega_k-\alpha\omega(i+\omega \tau))  & 0  
 \end{pmatrix}
 \begin{pmatrix}
 \delta m_x (k,\omega)\\
 \delta m_y  (k,\omega)\\
 \delta m_z  (k,\omega)
 \end{pmatrix}
 =-\begin{pmatrix}
  i\omega & 0 & 0 \\
 0 & i\omega & 0 \\
  0  & 0  & i\omega  
 \end{pmatrix}
 \begin{pmatrix}
 \delta m_x (k,\omega)\\
 \delta m_y  (k,\omega)\\
 \delta m_z  (k,\omega)
 \end{pmatrix}
 \]
  \begin{equation}\label{3x3matrixeqn}
 +\gamma\begin{pmatrix}
0\\
 -\mathcal{C}\left[H_{\mathrm{ac}}(k,\omega)\delta m_z  (k,\omega)\right]\\
  \mathcal{C}\left[H_{\mathrm{ac}}(k,\omega)\delta m_y  (k,\omega)\right]
 \end{pmatrix}
\end{equation}
\end{widetext}
where
\begin{equation}
    \omega_k=\gamma J M_s k^2
\end{equation}
and
\begin{equation}
    \alpha=\gamma M_s\eta
\end{equation}
with $\mathcal{C}\left(\mathbf{A}(\omega)\times \mathbf{B}(\omega)\right)$ representing the convolution of the vector functions $A$ and $B$ in frequency space. Taking the limit of vanishing external fields $H_0,H_{\mathrm{ac}}\rightarrow 0$, the above $3\times 3$ matrix equation can be written in a compact form
\begin{equation}
\label{dispersionrel}
\begin{array}{l}
\left[\mathrm{i}\omega\delta_{IJ}-( \mathrm{i}\gamma\eta\omega+\gamma\eta\tau\omega^2-\gamma\eta\tau v^2_\mathrm{iner}k^2  )\varepsilon_{IJK}M_{0,K}\right]=0
\end{array}
\end{equation}
where $I,J,K=x,y,z$ label the components of the vector $\mathbf{M}$ and $\sum_{K=x,y,z}M^2_{0,K}=M^2_s$. The above equations are subject to the constraint $\mathbf{M}_0\cdot\delta\mathbf{M}=0$, which from choosing $\mathbf{M}_0$ along $x$ gives $M_{0,K}=M_s\delta_{K,z}$ and reads in matrix form
\begin{equation}
\begin{pmatrix}
 M_s & 0 & 0 \\
 0 & 0 & 0 \\
  0  & 0  & 0  
 \end{pmatrix}
 \begin{pmatrix}
 \delta m_x (k,\omega)\\
 \delta m_y  (k,\omega)\\
 \delta m_z  (k,\omega)
 \end{pmatrix}=
 \begin{pmatrix}
 0\\
 0\\
 0
 \end{pmatrix}
\end{equation}
This constraint reduces the effective $3\times 3$ matrix in equation (\ref{3x3matrixeqn}) to an effective $2\times 2$ matrix equation, covering only the $y-z$ sectors of the original $3\times 3$ matrix equation, which can be written as
\[
 \begin{pmatrix}
 0 & -(\omega_k-\alpha\omega(i+\omega \tau)) \\
 (\omega_k-\alpha\omega(i+\omega \tau))  & 0 
 \end{pmatrix}\binom{\delta m_y(k,\omega)}{\delta m_z(k,\omega)}=
 \]
 \begin{equation}
 \begin{pmatrix}
  -i\omega & 0 \\
0  & -i\omega 
 \end{pmatrix}\binom{\delta m_y(k,\omega)}{\delta m_z(k,\omega)}
\end{equation}
The equation reduces to that for the determinant 
\begin{equation}
     \mathrm{det} \begin{pmatrix}
 i\omega & -(\omega_k-\alpha\omega(i+\omega \tau)) \\
 (\omega_k-\alpha\omega(i+\omega \tau))  & i\omega 
 \end{pmatrix}=0
\end{equation}
leading to the following fourth order polynomial equation for $\omega$
\begin{equation}\label{dispersionrelationequation}
    \omega^2=\left(\omega_k-\alpha\omega(i+\omega\tau)\right)^2
\end{equation}
Its solutions give the following dispersion relations 
\begin{equation}\label{dispersionrelations}
\omega=-\frac{\pm 1+i\alpha}{2\alpha\tau}\pm\frac{1}{2\alpha\tau}\sqrt{4\alpha\tau\omega_k+(\pm 1+i\alpha)^2}
\end{equation}
which are complex-valued in general, with imaginary part physically originates from the Gilbert damping term that leads to decay of the wave excitation and the associated dissipation of energy. The total four solutions consist of two pairs of two solutions; the two pairs differ only that they propagate in opposite directions. We will only describe the equations for the right-moving pair and rewrite its dispersions as
\begin{equation}\label{dispersionrelationsphysical}
\omega_{\pm}(k)=-\frac{i}{2\tau}\mp\frac{1}{2\alpha\tau}+\sqrt{v^2_{\mathrm{iner}}k^2+\frac{(1-\alpha^2)}{4\alpha^2\tau^2}\pm\frac{i}{2\alpha\tau^2}}
\end{equation}
\begin{equation}\label{dispersionrelationsphysicalmore}
\omega_{\pm}(k)=-\frac{i}{2\tau}\mp\frac{1}{2\alpha\tau}+
\Lambda_k(\cos\frac{\zeta}{2}\pm i\sin\frac{\zeta}{2})
\end{equation}
where
\begin{equation}
    \Lambda_k=\left[\left(v^2_{\mathrm{iner}}k^2+\frac{(1-\alpha^2)}{4\alpha^2\tau^2}\right)^2+\left(\frac{1}{2\alpha\tau^2}\right)^2\right]^{\frac{1}{4}}
\end{equation}
\begin{equation}
    \tan\zeta=\frac{\frac{1}{2\alpha\tau^2}}{v^2_{\mathrm{iner}}k^2+\frac{(1-\alpha^2)}{4\alpha^2\tau^2}}
\end{equation}
Writing explicitly the $\cos\frac{\zeta}{2}$ and $\sin\frac{\zeta}{2}$, we have
\begin{widetext}
\[
\omega_{\pm}(k)=\mp\frac{m}{2}+\sqrt{\frac{\sqrt{\left(v^2_{\mathrm{iner}}k^2+(\frac{m'}{2})^2\right)^2+\left(2\alpha(\frac{m}{2})^2\right)^2}+\left(v^2_{\mathrm{iner}}k^2+\left(\frac{m'}{2}\right)^2\right)}{2}}
\]
\begin{equation}\label{delicatequation}
+i\left(\pm\sqrt{\frac{\sqrt{\left(v^2_{\mathrm{iner}}k^2+(\frac{m'}{2})^2\right)^2+\left(2\alpha(\frac{m}{2})^2\right)^2}-\left(v^2_{\mathrm{iner}}k^2+\left(\frac{m'}{2}\right)^2\right)}{2}}-\frac{1}{2\tau}\right)
\end{equation}
\end{widetext}
where 
\begin{equation}
    m=\frac{1}{\alpha\tau}
\end{equation}
\begin{equation}\label{massrenorm}
m'=m\sqrt{1-\alpha^2}\simeq m
\end{equation}
where we have used (and we will continue doing so) the fact that $\alpha\ll 1$ in all physically realistic cases to simplify the expressions. In particular, for all values of wave vector $k$ applicable to this work, $v^2_{\mathrm{iner}}k^2+(m'/2)^2\gg 2\alpha(m/2)^2$, which allows us to simplify the expressions within the square roots in equation (\ref{delicatequation}) by expanding in $\alpha$ and leads to
\[
\omega_{\pm}(k)=\mp \frac{m}{2}+\sqrt{v^{2}_{\mathrm{iner}}k^2+\left(\frac{m}{2}\right)^2}
\]
\begin{equation}\label{simplifiedequation}
+i\left(\pm\alpha\left(\frac{m}{2}-\frac{v^2_{\mathrm{iner}}k^2}{m}\right)-\frac{1}{2\tau}\right)
\end{equation}
where we have immediately expanded out the imaginary part at small wave vector $k$ because we are merely interested in its $k$-dependence.

The real part of these physical solutions can thus be written as 
\begin{equation}\label{realdispersions}
\omega^R_{\pm}(k)=\mathrm{Re}[\omega_{\pm}(k)]=\mp\frac{m}{2}+\sqrt{v^2_{\mathrm{iner}}k^2+\left(\frac{m}{2}\right)^2}    
\end{equation}
The solution with upper index is gapless
\[
\epsilon_{\mathrm{sw}}(k)=\omega^R_{+}(k)=-\frac{m}{2}+\sqrt{v^2_{\mathrm{iner}}k^2+\left(\frac{m}{2}\right)^2}    
\]
\begin{equation}\label{spinwevdispersion}
\simeq \frac{v^2_{\mathrm{iner}}}{m}k^2=\gamma J M_s k^2 \end{equation}
corresponding to the spin wave dispersion at small wave vector $k$ which is gapless and non-relativistic (quadratic) in $k$.

The solution with lower index turns out to be gapped
\begin{equation}
\epsilon_{\mathrm{nw}}(k)=\omega^R_{-}(k)=\frac{m}{2}+\sqrt{v^2_{\mathrm{iner}}k^2+\left(\frac{m}{2}\right)^2}    
\end{equation}
indicating a cross over from a quadratic behavior to linear behavior at wave vector
\begin{equation}
    k_0=\frac{m}{2v_{\mathrm{iner}}}
\end{equation}
above which the nutation wave propagates with its 'speed of light' $v_{\mathrm{iner}}$. The resulting dispersion relations are illustrated in Fig. 3. in the main text. 

Writing $\omega_{\pm}(k)=\omega^R_{\pm}(k)-i\omega^I_{\pm}(k)$, the imaginary part $\omega^I_{-}(k)$ of the dispersion relation $\omega_{-}(k)$ describes the decay rate of the spin wave due to the Gilbert damping, given by
\begin{equation}
\omega^I_{+}(k)=\alpha\gamma J M_sk^2
\end{equation}
which indicates that the decay of the spin wave increases with the wave vector $k$. This suggests that the spin wave is robust only for really small wave vectors; $|k|\leq k_{\mathrm{sw}}$, which justifies the small $k$ expansion that leads to equation (\ref{spinwevdispersion}). Comparing with experiment suggests that $k_{\mathrm{sw}}\simeq 0.14 k_0$ as stated in the main text. On the other hand, the decay rate of the nutation wave is given by
\begin{equation}
\omega^I_{-}(k)=\frac{1}{\tau}-\alpha\gamma J M_sk^2
\end{equation}
decreases with the wave vector $k$ until
\begin{equation}
    k_{\mathrm{UV}}=\sqrt{\frac{1}{\alpha\tau\gamma JM_s}}=\frac{m}{v_{\mathrm{iner}}}=2k_0
\end{equation}
when the nutation wave starts to grow in its amplitude, marking the breakdown of the theory. We can define the characteristic decay time of the nutation wave
\begin{equation}
    \tau_d=\frac{2\pi}{\omega^I_{-}(k_0)}=\frac{8\pi\tau}{3}
\end{equation}
and the corresponding decay length
\begin{equation}
    l_d=v_{\mathrm{iner}}\tau_d=\frac{8\pi v_{\mathrm{iner}}\tau}{3}
\end{equation}

Now, we evaluate the above quantities using some real material parameters (e.g. $\alpha=0.1, a=2 \AA,M_s=10^6$A/m, $J=J_{\mathrm{exc}}/(M^2_sa),J_{\mathrm{exc}}=40$K,$\gamma=1.75\times 10^{11}$sA/kg) that we discussed in the main text. We obtain $\tau_d=(8\pi/3)\times 10^{-14}$s$\gg \tau=10^{-14}s$ which implies a relatively slow decay rate or relatively large decay time if we note that within the time scales $t\lesssim \tau=10^{-14}s$ of interest, the nutation wave decay will not be observable and the wave thus exists as a robust excitation. This also holds for larger choice of $\tau$, e.g. $\tau=10^{-12}s$. The more detailed robustness analysis is as given in the main text.

The characteristic speed of the nutation wave at large wave vector $k\gtrsim k_0$ is given by $v_{\mathrm{iner}}$, which is found to be superior to that of the spin wave in most cases. Using the above realistic material parameters for example, we find that $v_{\mathrm{iner}}=22,000$m/s compared to typical spin wave speed $v_{\mathrm{sw}}\approx 3,000$m/s. This also holds for larger choices of $\tau$, e.g. $\tau=10^{-12}s$ if we have smaller but yet realistic Gilbert damping constant, e.g. $\alpha=0.01$, which gives $\tau_d=(8\pi/3)\times 10^{-12}$s $> \tau=10^{-12}$s and $v_{\mathrm{iner}}=7,000$m/s. What is most interesting from the above analyses is that the decay rate of the nutation wave is finite at zero wave vector but it decreases with the wave vector. Therefore, nutation wave with shorter wave lengths is actually more robust than longer ones, up until a critical wave length $\lambda=2\pi/k_{\mathrm{UV}}$. This is perfect for our intended purpose because at larger wave vectors, the nutation wave enters its linear regime for which the nutation wave reaches its optimum speed $v_{\mathrm{iner}}$, thus enforcing the superiority of the nutation wave over the spin wave in terms of their speed and frequency. 

The group velocity corresponding to the relativistic dispersion relation of the nutation wave is given by
\begin{equation}\label{groupvelocity}
v_g(k)=\frac{\partial \epsilon_{\mathrm{nw}}(k)}{\partial k}=\frac{v^{'2}_{\mathrm{iner}}k}{\sqrt{v^{'2}_{\mathrm{iner}}k^2+m'^2}}
\end{equation}
which suggests that $v_g(k)\simeq v'_{\mathrm{iner}}$ at large enough $k$. The averaged group velocity in Eq.(8) can be defined by
\begin{equation}\label{averagegroupvelocity}
v_g=\frac{\int^{\infty}_0 dk f(k) n(k)v_g(k)}{\int^{\infty}_0 dk f(k) n(k)}
\end{equation}
where $n(k)$ is the density of states and $f(k)=(\exp(\epsilon(k)/T)-1)^{-1}$ is the Bose-Einstein distribution function for the nutaton. We make the simplest approximation of constant density of states $n(k)=L/(2\pi)$ which implies that we have one nutation state per unit cell in the reciprocal $k$ space. Since we are considering a one-dimensional spin system which can have a broken-symmetry state at $T=0$, we will have to take the $T\rightarrow 0$ limit at the end. We also have to note that since the nutatons are bosons, at $T=0$ we will have a Bose-Einstein condensation of nutatons where they will lump together at the lowest-energy state at $k=0$ with energy $\epsilon_{\mathrm{nw}}(k=0)=m$. Evaluating the integrals in the numerator and denominator of Eq.(\ref{averagegroupvelocity}) gives 
\begin{equation}\label{averagegroupvelocityevaluation}
v_g=\frac{\left[m-\sqrt{m^2+\frac{v^2_{\mathrm{iner}}\pi^2}{a^2}}+T\log\left(\frac{1-e^{\frac{\sqrt{m^2+\frac{v^2_{\mathrm{iner}}\pi^2}{a^2}}}{T}}}{1-e^{\frac{m}{T}}}\right)\right]}{N(k=0)+\frac{e^{-\frac{m}{T}} \sqrt{m} \sqrt{\frac{\pi}{2}} \sqrt{T}
  \mathrm{Erf}[\frac{\pi v_{\mathrm{iner}}}{\sqrt{2} a \sqrt{m} \sqrt{T}}]}{v_{\mathrm{iner}}}}
\end{equation}
where we have imposed an ultraviolet cutoff $\Lambda=\pi/a$ where $a$ is the lattice spacing for the upper limit of the integral over the wave vector $k$ and $L$ is the system size and we have neglected the renormalization effect of $\omega_n$ and $\tau$ by taking $v'_{\mathrm{iner}}=v_{\mathrm{iner}}, m'=m$. We take the number of condensed bosons $N(k=0)$ to be the number of lattice sites (unit cells) $N(k=0)=L/a$. The remaining terms in the denominator of Eq.(\ref{averagegroupvelocityevaluation}) vanish in the limit $T\rightarrow 0$ as expected. The numerator also vanishes at $T=0$ due to the fact that all the bosons condense at $k=0$ at which the group velocity is zero, according to Eq.(\ref{groupvelocity}). As a result, $v_g=0$. However, if we apply an electromagnetic wave to excite the nutatons, some of the condensed nutatons will acquire the momentum transferred by the magnetic field part of the wave and attain finite $k$'s at higher energies which leads to
\begin{equation}\label{averagevg}
v_g=K v_{\mathrm{iner}}
\end{equation}
where $0<K\leq 1$. With $v_g=K v_{\mathrm{iner}}$, we obtain 
\begin{equation}
\tilde{v}_{s=\pm}=s v_{\mathrm{iner}} \pm v_g=v_{\mathrm{iner}} (s\pm K)
\end{equation}
Eq.(8) in the main text already possesses a space-time translational invariance for arbitrary values of $0<K\leq 1$. At a special point when $s\pm K=\pm 1/2$ such that all the velocity variables appearing in Eq.(8) in the main text become $|\tilde{v}_g|,|\tilde{v}_{s=\pm}|=v_{\mathrm{iner}}/2$, we get a full Lorentz invariance. One can think of this special point as a resonance state where the whole system moves as if it were a rigid body described by a wave of unique velocity $|\tilde{v}_g|,|\tilde{v}_{s=\pm}|=v_{\mathrm{iner}}/2$. It is important to note that if we consider a three-dimensional ferromagnetic spin system instead, we can go up at finite $T$ and the average group velocity will naturally be nonzero as a nonzero $K$ can be achieved even without an application of electromagnetic (e.g, optical) wave. That is, an ordinary AC magnetic field suffices to generate the propagating nutation wave. 

\section{Numerical Studies}
The explicit solution of the LLG equation with the inertia and exchange field term Eq.(5) in the main text can be determined  most usefully by numerical integration. We rewrite Eq.(5) in terms of the spherical angles $\theta,\phi$ starting with $\mathbf{M}(x,t)=M_s\hat{\mathbf{e}}_r$ where $M_s$ is fixed while the radial unit vector $\hat{\mathbf{e}}_r=\hat{\mathbf{e}}_r(x,t)$ is itself function of $x$ and $t$. Following the derivation given in \cite{APLs} and \cite{JAPs}, we obtain
\begin{widetext}
\begin{equation}\label{EOMtheta}
\ddot{\theta}-\frac{J}{\eta\tau}\theta^{''}=-\frac{\dot{\theta}}{\tau}+\left({\dot{\phi}}^2\cos\theta-\frac{\dot{\phi}}{\tau_1}-\frac{\omega_2}{\tau_1}\right)\sin\theta+\frac{\omega_{\mathrm{AC}}(t)}{\tau_1}\cos\theta\cos\phi-\frac{J}{\eta\tau}\frac{{\phi '}^2\sin 2\theta}{2},
\end{equation}
\begin{equation}\label{EOMphi}
\ddot{\phi}-\frac{J}{\eta\tau}\phi^{''}=\frac{1}{\sin\theta}\left[\frac{\dot{\theta}}{\tau_1}-\frac{\dot{\phi}\sin\theta}{\tau}-2\dot{\theta}\dot{\phi}\cos\theta-\frac{\omega_{\mathrm{AC}}(t)}{\tau_1}\sin\phi+2\frac{J}{\eta\tau}\theta'\phi'\cos\theta\right]
\end{equation} 
\end{widetext}
where $\alpha=\gamma\eta M_s, \tau_1=\alpha\tau, \omega_2=\gamma H, \omega_{\mathrm{AC}(t)}=\gamma H_{\mathrm{AC}}(t)$ are respectively the dimensionless Gilbert damping, the relaxation time, the DC and AC Larmor frequencies, while $\theta=\theta(x,t),\phi=\phi(x,t),\theta'=\partial\theta/\partial x,\phi'=\partial\phi/\partial x,\dot{\theta}=\partial\theta/\partial t,\dot{\phi}=\partial\phi/\partial t, $ etc. We have chosen the DC magnetic field to be along $z$ axis and AC magnetic field of the electromagnetic wave to be polarized along $x$ axis. The exchange field gives rise to the appearance of spatial derivative terms of $\theta$ and $\phi$ in Eqs.(\ref{EOMtheta}-\ref{EOMphi}). 

The results of the numerical solution  of the above equations are presented in Fig. 1. in the main text. As we can see there, each of the spins oscillates with time and as one can see already, immediately after $t=0$, a roughly periodic spatial variation develops (appearing like the teeth of a saw) and which travels across the chain. This is the signature of the nutation wave that we have earlier concluded analytically. The wave is periodic in space but not sinusoidal which means that it can be decomposed into different Fourier wave vector components $k$. The amplitude of the wave quickly decays over time due to the damping. 

In solving numerically the solution to Eqs.(\ref{EOMtheta}-\ref{EOMphi}), we have set the initial conditions at $t=0$ such that the orientation of the spins defines a sinusoidal function of $x$. We do not set any final conditions nor boundary conditions since the dynamics involve damping so that while there may be periodicity in space and time, the amplitudes decay with time. The justification for this open boundary condition is that while the system is deterministic (i.e. not stochastic), the dynamics is dissipative; i.e. it does not preserve the energy. On the other hand, in order to describe a non--trivial gauge field, we choose the boundary conditions in such a way that the phase of the complex scalar $\Phi=M_x+\mathrm{i}M_\perp$ changes sign from one end of the chain to the other. 

To produce Fig. 1 in the main text, we have used the following parameter values: 
\[
v_{\mathrm{iner}}=22,200 \mathrm{m/s}, \alpha=0.1,\tau=10^{-14}\mathrm{s}, a=2\times 10^{-10}\mathrm{m}
\]
\[
H_0=0, H_{ac}=1 T, \omega_{ac}=2\pi \times 10^{14}\mathrm{rad/s}
\]
where the AC magnetic field of the linearly-polarized electromagnetic wave is given by $H_{ac}(t)=H_0+H_{ac}\sin\omega_{ac}t$. The $\tau$ sets the time scale while the $a$ sets the length scale. We have used as the initial condition $\phi(x,t=0)=0.01 \sin kx, \theta(x,t=0)=\pi/2+0.01\sin kx$, which gives rise to a propagating wave as shown in Fig.1. in the main text.

We experimented numerically by setting $J=0$ and found that the wave profile shown in Fig. 1 in the main text vanishes completely as shown in the figures below.
\begin{figure}
 \includegraphics[angle=0,origin=c, scale=0.30]{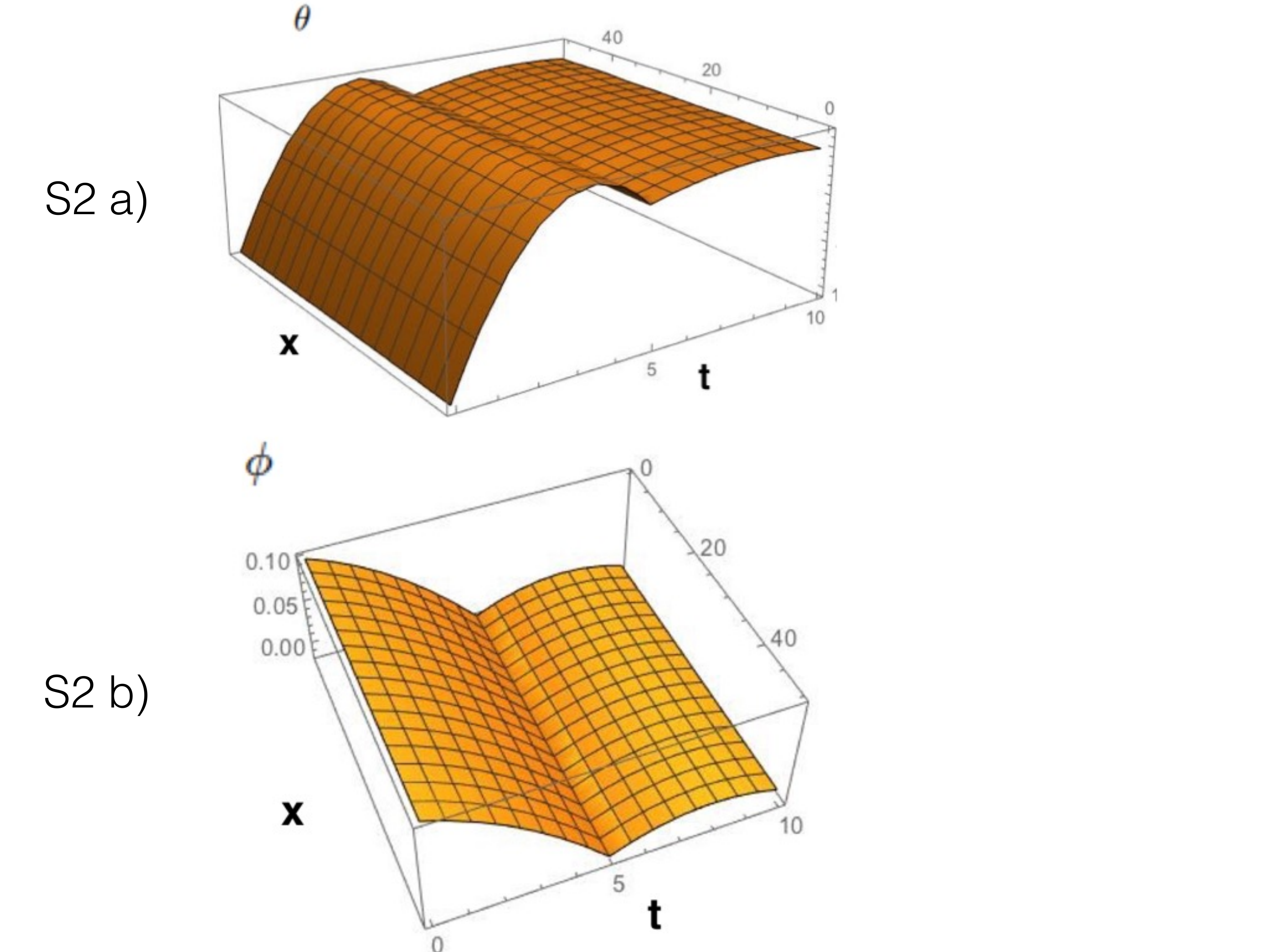}
 \label{fig:NutatonWaveProfile}
 \caption{
 The evolution of the spherical angles of the spins on a spin chain $\theta(x,t)$(left/top) and $\phi(x,t)$(right/bottom) for $J=0$.}
 \end{figure}
On the other hand, when we set $H_{ac}=0$, the wave profile still exists, but with reduced amplitude. This simply suggests that the wave is due to  the exchange interaction and not by the excitation by the AC magnetic field of the electromagnetic wave. In fact the exchange interaction acts as the mediator of the nutation wave while the AC field excites the wave. This is very analogous to the fact that electromagnetic interaction mediates the propagation of electromagnetic wave and that an AC electric source is needed to excite the electromagnetic wave. We also tried to use a spatially uniform initial condition, for example, $\phi(x,t=0)=0, \phi(x,t=0)=\pi/2$ and found that there is a spatial periodic pattern along the spin chain due to the fact that the uniform state is not an eigenstate of the system, but the periodic modulation does not propagate as the time goes and decays due to the Gilbert damping. 

\section{The Strength of the  Nutation Wave}
With regard to the excitation of nutation wave by an electromagnetic wave, we obtain from the linear analysis Eq.(\ref{linearanalysis}) the strength of the wave,
\begin{equation}
|\tilde{\mathbf{M}}(k,\omega)|=\frac{i}{\omega}\gamma |\mathbf{M}_0||\mathbf{H}_{\mathrm{AC}}(\omega)|e^{i\Delta\psi_{\mathbf{H}_{\mathrm{AC}},\tilde{\mathbf{M}}}}\sin\Delta\phi_{\mathbf{H}_{\mathbf{M}_0,\mathrm{AC}}}
\end{equation}
where $\Delta\psi_{\mathbf{H}_{\mathrm{AC}},\tilde{\mathbf{M}}}=\psi_{\mathbf{H}_{\mathrm{AC}}(\omega)}-\psi_{\tilde{\mathbf{M}}}$ is the complex phase angle difference and $\Delta\phi_{\mathbf{H}_{\mathrm{AC}}}=\phi_{\mathbf{H}_{\mathrm{AC}}}-\phi_{\mathbf{M}_0}$ is the difference in the azimuthal angle between $\mathbf{M}_0$ and $\mathbf{H}_{\mathrm{AC}}$. Note that $\tilde{\mathbf{M}}(k,\omega)$ is a complex-valued O(3) vector and the $|\cdots|$ represents the length of such vector in the vector space (not in the complex plane space), which is thus still complex-valued in general. This result suggests that the magnitude of nutation is proportional to the magnitude of $\mathbf{M}_0$, the amplitude of the electromagnetic driving field $\mathbf{H}_{\mathrm{AC}}$ and the angle between them and decreases with the electromagnetic wave frequency $\omega$. The inverse proportionality on the frequency is in perfect agreement with the same behavior for the transverse susceptibility of the nutation peak found in \cite{JAPs}. One should remember that this amplitude preserves the magnitude of the magnetization $M_s$. 

\section{Dynamical Inertia vs Topological Inertia}
It is interesting to compare the inertia term that arises in short time scale of magnetization dynamics considered in this work with the one that arises in topological spin texture such as magnetic skyrmions. In our work, the spin texture considered is just a uniform (or nearly uniform) ferromagnet with no topological defect or texture characterized by a topological invariant. The spin Berry phase plays no role in the dynamics of the magnetization under consideration and the inertia term that arises relates to high energy excitations. In contrast, the inertia term that appears in magnetic skyrmion originates from the Berry phase effect giving rise to topological invariant that acts as a magnetic charge and that leads to gyrotropic motion and chiral modes where the chirality arises from the superposition of two modes propagating with opposite but different velocities \cite{MakhfudzPRL2012s}. No modes of chiral nature exist in the problem that we study, in accordance with the nutation wave dispersion which is completely symmetric just like that of the standard spin wave, as shown in Fig. 3. 

\section{Note on Nutation Wave Dispersion Relation}
According to relativistic theory \cite{MondalPRB2017s}, the magnetic moment inertia is given by 
\begin{equation}\label{magneticinertia}
\mathcal{I}_{ij}=\frac{\zeta\hbar}{2mc^2}\left[\mathbb{I}+\mathrm{Re}(\chi^{-1}_m)_{ij}\right]
\end{equation}
where $\zeta=\mu_0\gamma \hbar/(4mc^2)$. Since $\mathcal{I}\sim \gamma \eta\tau$, we thus conclude that $\omega_n\sim mc^2$ where $m$ is the mass of the fermion and $c$ is the speed of light. The nutation wave gap is thus of the order of the rest mass energy of the fermion. If we take this fermion to be the electron and neglect the contribution of $\chi_m$ in Eq.(\ref{magneticinertia}), we obtain $\omega_n$ of the order of $10^{21}$ rad s$^{-1}$, which is an extremely high frequency, while typical nutation resonance frequency is on the order of $10^{15}$ rad s$^{-1}$ \cite{AtomisticS}. Certainly this suggests rather an estimate for the upper bound to the nutation frequency and that the renormalization effect of $\chi_m$ in Eq.(\ref{magneticinertia}) should be properly taken into account. 

As a final note, while we have put aside the electronic degree of freedom from the picture, our theory applies to ferromagnetic metals that are used in data storage devices, where the magnetization is produced by the localized ($d$ or $f$-orbital) electron spins. New effects due to the presence of the $s$-conduction electrons and their coupling to localized electrons are open questions that we will address for a future work.


\begin{thebibliography}{1}

\bibitem{RMPspintronics}A. Fert, Rev. Mod. Phys. 80, 1517 (2008).

\bibitem{UltrafastReview1}E. Beaurepaire, J.-C. Merle, A. Daunois, and J.-Y. Bigot, Phys. Rev. Lett. 76, 4250 (1996).

\bibitem{UltrafastReview2}Th. Gerrits, H. A. M. van den Berg, J. Hohlfeld, L. B\:{a}r and Th. Rasing, Nature 418, 509 (2002).

\bibitem{RMPultrafast}A. Kirilyuk, A. V. Kimel, and T. Rasing, Rev. Mod. Phys. 82, 2731 (2010).

\bibitem{Ultimate}Tudosa, I., C. Stamm, A. B. Kashuba, F. King, H. C. Siegmann, J. St\:{o}hr, G. Ju, B. Lu, and D. Weller, 2004, Nature London 428, 831.

\bibitem{LL}L. Landau and E. Lifshitz, Phys. Z. Sowjetunion 8, 153 (1935).

\bibitem{GilbertIEEE}T. L. Gilbert, IEEE Trans. Mag. 40, 3443 (2004).

\bibitem{Ciornei}M.-C. Ciornei, J. M. Rubi, and J.-E. Wegrowe, Phys. Rev. B 83, 020410(R) (2011).

\bibitem{AJP}J.-E. Wegrowe and C. Ciornei, Am. J. Phys. 80, 607 (2012).

\bibitem{APL}E. Olive, Y. Lansac, and J.-E. Wegrowe, Appl. Phys. Lett. 100, 192407 (2012).

\bibitem{JAP}E. Olive, Y. Lansac, M. Meyer, M. Hayoun, and J.-E. Wegrowe, J. Appl. Phys. 117, 213904 (2015).

\bibitem{FermiSurfaceBreathingModel}M. F\:{a}hnle, D. Steiauf, and C. Illg, Phys. Rev. B 84, 172403 (2011).

\bibitem{Atomistic}S. Bhattacharjee, L. Nordstr\:{o}m, and J. Fransson, Phys. Rev. Lett. 108, 057204 (2012).

\bibitem{MondalPRB2017}R. Mondal, M. Berritta, A. K. Nandy, and P. M. Oppeneer, Phys. Rev. B 96, 024425 (2017).

\bibitem{MakhfudzPRL2012}I. Makhfudz, B. Kr$\ddot{\mathrm{u}}$ger, and O. Tchernyshyov, Phys. Rev. Lett. 109, 217201 (2012).

\bibitem{Neeraj}K. Neeraj, N. Awari, S. Kovalev, D. Polley, N. Z. Hagström, S. S. P. K. Arekapudi, A. Semisalova, K. Lenz, B. Green, J-C. Deinert, \textit{et al.}, arXiv:1910.11284 (2019).

\bibitem{Bastardis}R. Bastardis, F. Vernay and H. Kachkachi, Phys. Rev. B 98, 165444 (2018).

\bibitem{LandauLifshitzStatPhys2}L. D. Landau and E. M. Lifshitz,  \textit{Statistical Physics} Part 2 (Pergamon Press, New York, 1980).

\bibitem{WikipageAlembert}$\mathrm{https://en.wikipedia.org/wiki/Wave\_equation}$

\bibitem{Auerbach}A. Auerbach, \textit{Interacting Electrons and Quantum Magnetism} (Springer-Verlag New York, Inc. (1994)).

\bibitem{HeisenbergFerromagnetSpinchain1}R. D. Willett, C. P. Landee, and D. D. Swank: J. Appl. Phys. 49 (1978) 1329.

\bibitem{HeisenbergFerromagnetSpinchain2}D. D. Swank, C. P. Landee, and R.
D. Willett, Phys. Rev. B 20, 2154 (1979).

\bibitem{HeisenbergFerromagnetSpinchain3}R. D. Willett, C. P. Landee, R. M. Gaura, D. D. Swank, H. A. Groenendijk, and A. J. van Duynevelt: J. Magn. and Magn. Mater. 15 (1980) 1055.

\bibitem{NatureNano}M. Madami, S. Bonetti, G. Consolo, S. Tacchi, G. Carlotti, G. Gubbiotti, F. B. Mancoff, M. A. Yar, and J. Åkerman, Nature Nanotechnology 6, 635 (2011).

\bibitem{NatComms}C. Liu, J. Chen, T. Liu, F. Heimbach, H. Yu, Y. Xiao, J. Hu, M. Liu, H. Chang, T. Stueckler, \textit{et al.}, Nat. Commun. 9, 738 (2018).

\bibitem{Higgs1}P. W. Higgs, Phys. Rev. Lett. 13, 508 (1964).

\bibitem{Higgs2}F. Englert and R. Brout, Phys. Rev. Lett. 13, 321 (1964).

\bibitem{Sherman}D. Sherman, U. S. Pracht, B. Gorshunov, S. Poran, J. Jesudasan, M. Chand, P. Raychaudhuri, M. Swanson, N. Trivedi, A. Auerbach, \textit{et al.}, Nat. Phys. 11, 188 (2015).

\bibitem{Wessel}M. Loh\:{o}fer and S. Wessel, Phys. Rev. Lett. 118, 147206 (2017).

\bibitem{Sachdev}S. Sachdev, \textit{Quantum Phase Transitions} (Cambridge University Press, Cambridge, UK, 2011).

\bibitem{PekkerVarma}D. Pekker and C.M. Varma, Annual Review of Condensed Matter Physics 2015 6:1, 269-297.

\bibitem{PeskinSchroederQFT}M. Peskin and D. Schroeder, \textit{Introduction to Quantum Field Theory} (Perseus, Cambridge, MA, 1995).
 
\bibitem{MakhfudzPRB2014}I. Makhfudz, Phys. Rev. B 89, 024401 (2014).

\bibitem{MakhfudzJPCM}I. Makhfudz, Journal of Physics: Condensed Matter 30, 225801 (2018).

\bibitem{Mannheim}P. D. Mannheim, arXiv:1506.04120v2 (2015).

\bibitem{nussle2019dynamic} Th. Nussle, P.  Thibaudeau and S. Nicolis, Phys. Rev. B100, 214428 (2019).


\end{thebibliography}

\begin{thebibliography}{1}

\bibitem{APLs}E. Olive, Y. Lansac, and J.-E. Wegrowe, Appl. Phys. Lett. 100, 192407 (2012).

\bibitem{JAPs}E. Olive, Y. Lansac, M. Meyer, M. Hayoun, and J.-E. Wegrowe, J. Appl. Phys. 117, 213904 (2015).

\bibitem{AtomisticS}S. Bhattacharjee, L. Nordstr\:{o}m, and J. Fransson,Phys. Rev. Lett. 108, 057204 (2012).

\bibitem{MakhfudzPRL2012s}I. Makhfudz, B. Kr$\ddot{\mathrm{u}}$ger, and O. Tchernyshyov, Phys. Rev. Lett. 109, 217201 (2012).

\bibitem{MondalPRB2017s}R. Mondal, M. Berritta, A. K. Nandy, and P. M. Oppeneer, Phys. Rev. B 96, 024425 (2017).

\end{thebibliography}
\end{document}